\documentclass{aa}  
\usepackage[varg]{txfonts}
\usepackage{graphicx}
 \usepackage{lscape}
\usepackage{txfonts}
\begin{document}

   \title{Quasi-periodicities of BL Lac objects
  }
\subtitle{}

       \author{A. Sandrinelli \inst{1,2}, 
                  S. Covino \inst{2},
                  A. Treves \inst{1,2},
                  A. M. Holgado \inst{3},
                  A. Sesana \inst{4},
                  E. Lindfors \inst{5}, 
                  V.  F. Ramazani  \inst{5}
                        }

   \institute{Universit\`a dell'Insubria, Dipartimento di Scienza ed Alta Tecnologia,
              Via Valleggio 11, I-22100, Como, Italy      
              \and
            INAF-Istituto Nazionale di Astrofisica, Osservatorio Astronomico di Brera, 
             Via Bianchi 46, I-23807 Merate (LC), Italy      
            \and 
            Department of Astronomy and National Center for Supercomputing Applications, 
             University of Illinois at Urbana-Champaign, Urbana IL, 61801, USA        
             \and 
             School of Physics and Astronomy and Institute of Gravitational Wave Astronomy, 
             University of Birmingham, Edgbaston B15 2TT, United Kingdom       
            \and 
             Tuorla Observatory, Department of Physics and Astronomy, University of Turku, Finland               
            \\
            \email{asandrinelli@yahoo.it}
              }
              
   \date{Received ....; accepted ....}

% \abstract{}{}{}{}{} 
% 5 {} token are mandatory%   \title{Quasiperiodicities of BL Lac Objects and Their Origin\\
%\today}

  \abstract
  % context heading (optional)
  % {} leave it empty if necessary  
{We review the reports of possible year-long quasi-periodicities of BL Lac objects 
in the $\gamma$-ray and optical bands, and  present a homogeneous 
  time analysis of the light curves of PKS2155$-$304, PG1553+113, and BL Lac.  
Based on results from a survey covering the entire  \textit{Fermi} $\gamma$-ray sky
we have estimated the fraction  of possible quasi-periodic BL Lac objects.
 We compared the cyclical behaviour in BL Lac objects with that derived from 
 the search of possible optical periodicities in quasars, 
and find that at z$\lesssim$1 the cosmic density of quasi-periodic BL Lac objects
 is larger than that of quasi-periodic quasars.
 If the BL Lac quasi-periodicities
 were due to a supermassive binary black hole (SBBH) 
  scenario, there could be a tension with the upper limits on the gravitational wave background
   measured by the pulsar timing array. 
 The argument clearly indicates the difficulties of generally 
   associating quasi-periodicities of BL Lac objects with SBBHs.
   }


\titlerunning{Quasi-periodicities of BL Lac Objects and Their Origin}
\authorrunning{A. Sandrinelli, S. Covino, A. Treves}
 \maketitle

%________________________________________________________________

\section{Introduction}\label{theintro}

Objects known as BL Lac are active galactic nuclei characterized by 
high variability, large polarization, and weakness of emission lines
 \citep[e.g.][]{Falomo2014,Madejski2016,Padovani2017}.
They are generally interpreted as systems in which a relativistic jet
is pointing towards the observer direction. 
The jet emission dominates over other components, such as, for example, 
the thermal flux of the accretion disk. Their variability  is observed 
at all wavelengths, and on timescales ranging  from a few 
seconds to years. Searches for quasi-periodicities initially focused
on the optical and X-ray bands yielding controversial results,
the best known being the proposal of a 12 year periodicity for OJ 287
\citep[][]{Sillanpaa1988,Lehto1996}.
The situation has evolved in the last decade, because of the 
continuous monitoring of the $\gamma$-ray sky by the \textit{Fermi} mission, 
which demonstrated that BL Lac objects are the dominant component 
of the extragalactic sky, and was able to construct light curves 
of the brightest BL Lacs, which have typical integration times of weeks. 
The optical monitoring of BL Lacs was also much increased, 
as a result of the diffusion of robotic telescopes.

In 2014 we began a search for periodicities in the \textit{Fermi} $\gamma$-ray 
light curves of bright BL Lac objects. We found some evidence \citep{Sandrinelli2014a}
 of a   quasi-periodic oscillation of  T$\sim$630 days in PKS2155$-$304
  (z=0.116), corresponding to twice the optical
 period originally proposed by \cite{Zhang2014}. 
 The indication of   both   $\gamma$-ray and optical periods was confirmed by 
 \cite{Sandrinelli2016a}. A further confirmation of the $\gamma$-ray period came from \cite{Zhang2017a}. 
 \cite{Ackermann2015} provided evidence of a quasi-periodicity of 2.18 yr in the
 \textit{Fermi} light  curve of PG1553+113  (z$\sim$0.4), with counterparts in
  the optical and radio. 
We found  \citep{Sandrinelli2016b,Sandrinelli2017} year-long possible $\gamma$-ray
and optical correlated quasi-periodicities in PKS0537$-$441 (z=0.892)
 and BL Lac  (z=0.069), corresponding to 0.77 yr and 1.86 yr, respectively. 
 Recently \cite{Zhang2017b}  detected a 2.1 yr $\gamma$-ray quasi-periodic 
 oscillation in PKS0301$-$243 (z=0.260).
Most of these results are at a modest statistical significance, yet taken together 
they represent an interesting sample.

 \begin{figure*}
\centering
\includegraphics[trim=0.5cm 2.5cm 0.cm 12.5cm clip=true,width=2\columnwidth]{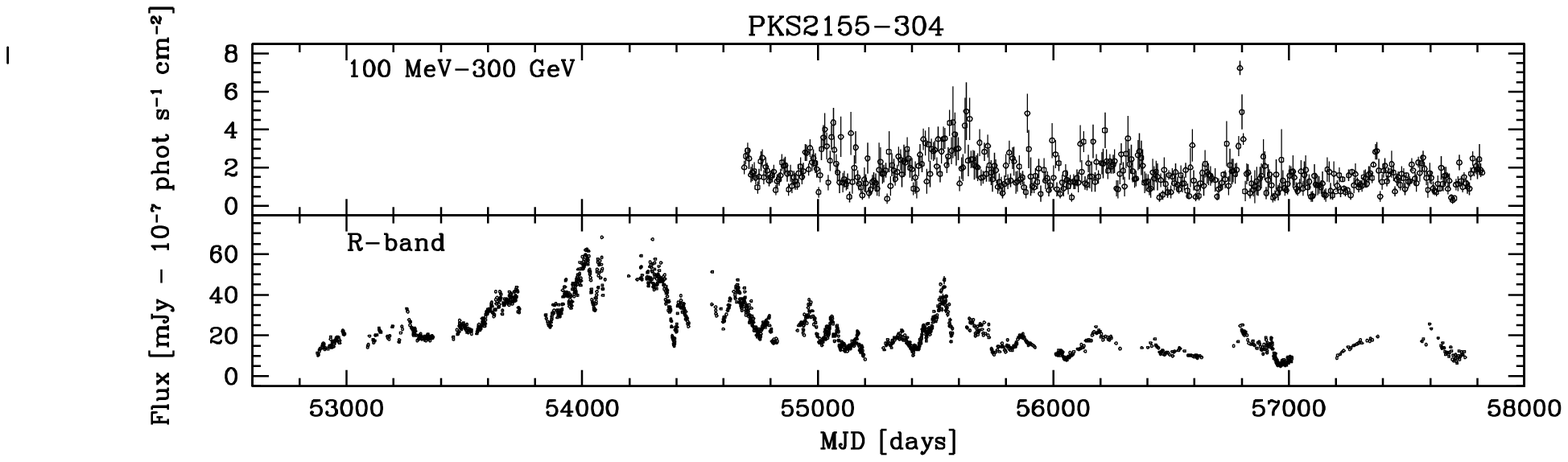} 
\includegraphics[trim=0.5cm 2.cm 0.cm 12cm, clip=true,width=2\columnwidth]{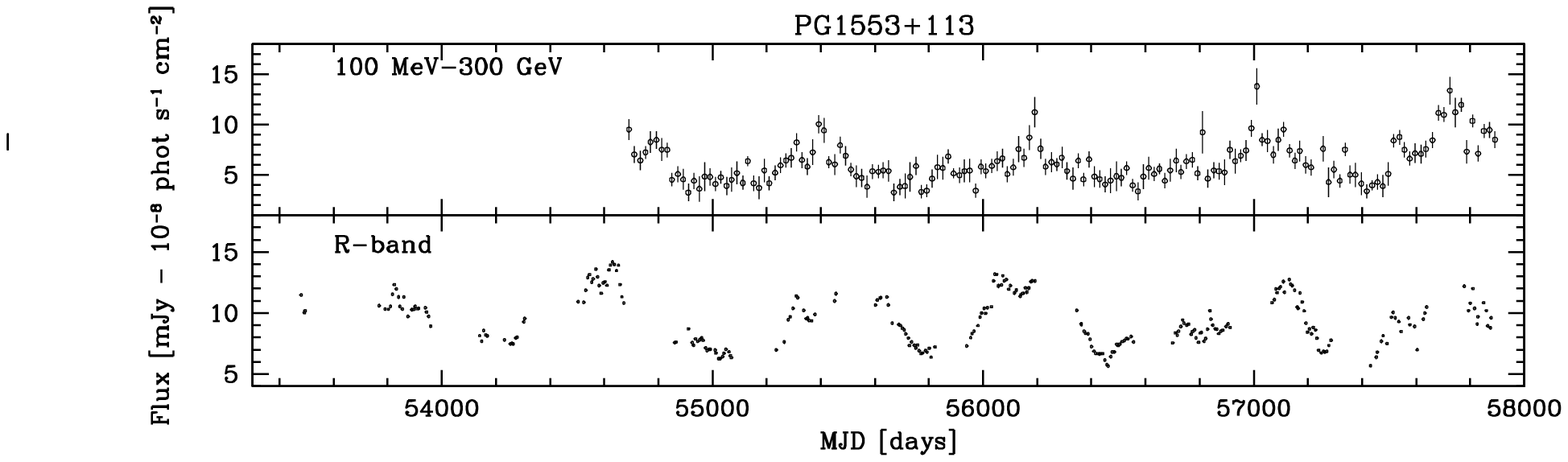} 
\caption{ \label{lc} 
Flux light curves of PKS2155$-$304 (\textit{top panel}) and PG1553$+$113 (\textit{bottom panel}).
 $Fermi$ light curves in the 100 MeV-300 GeV energy range  (7-day and 20-day binned, respectively)
  are reported. 
 The combined nightly averaged R-band data are also  given.  
The light curves are not host galaxy subtracted.
In the optical light curves error bars are in most cases smaller than symbol size. 
} 
\end{figure*}

Different methods were used for the search of quasi-periodicities, 
and for the estimation of their statistical significance. 
 In this paper (Section \ref{thesearch}) we concentrate on the procedure originally
described by  \citet[][see also \citeauthor{Guidorzi2016}
 \citeyear{Guidorzi2016}]{Vaughan2005,Vaughan2010}, 
which we have already applied  to the case of BL Lac \citep{Sandrinelli2017}.  
We focus on the $\gamma$-ray and optical light curves of PKS2155$-$304, 
and PG1553+113, in order to achieve a homogeneous analysis of three sources.

As already noted in \cite{Covino2017} and \cite{Sandrinelli2017},
 indications of year-long quasi-periodicities appear much more frequently in BL Lac objects 
 than in quasars, and in our discussion (Section \ref{thediscussion}) we refer,
  in particular,  to recent results 
 \citep{Sesana2018}, which  in the case of quasars may constrain the interpretation of 
 quasi-periodicities as being due to the binary period of a pair of supermassive black holes. 
 The argument is based on the upper limits of the gravitational wave background.  
 We then compare the local densities of quasi-periodic quasars, and BL Lac objects, 
 and concentrate on the interpretation of quasi-periodicities of the latter class.

%__________________________________________________________________

\section{\label{thesearch} The search of quasi-periodicities in PKS2155$-$304 and PG1553+113}

%______________________________________________ 
\subsection{\label{thelc} Light curves}

In Figure, \ref{lc} we report the optical and $\gamma$-ray light 
curves of PKS2155-304 and of PG1553+113  (100 MeV - 300 GeV, R band).
For  PKS2155$-$304 the $\gamma$-ray light curve was taken from the \textit{Fermi}
site\footnote{\texttt{http://fermi.gsfc.nasa.gov/ssc/data/access/lat/msl\_lc/\label{fermi}}} 
with a one week integration bin.
  Nightly averaged  data points were derived in the  R optical band from 
  the Rapid Eye Mounting Telescope photometry \citep[REM\footnote{ 
\texttt{http://www.rem.inaf.it}},][]{Zerbi2004, Covino2004,Sandrinelli2014b},  
and combined with data  drawn from the Small \& Moderate Aperture
 Research Telescope System archives
  \citep[SMARTS\footnote{\texttt{http://www.astro.yale.edu/smarts/glast/home.php}},][]{Bonning2012}, 
  the Tuorla Blazar Monitoring Program{\footnote{\texttt{http://users.utu.fi/kani/1m}}}  \citep{Takalo2008},
  the  Steward Observatory Fermi Blazar Observational 
  Program\footnote{\texttt{http://james.as.arizona.edu/$\sim$psmith/Fermi/}} 
  \citep{Smith2009}, and with observations from ROTSE-III and the  All Sky Automated Survey 
robotic telescopes and archival data  collected by \cite{Kastendieck2011},
  securing almost regular coverage.
   For PG1553+113 the R-light curve is mainly from Tuorla 
 Observatory\footnotemark[4]{} 
  with some integration from REM\footnotemark[2]{}. 
The $\gamma$-ray curve, with a binning of 20 days, is from \cite{Ackermann2015}
   and \cite{Cutini2016},  and complemented with data from the  \textit{Fermi}
    site\footnotemark[1]{}. 
 
%----------------------------------------------------

\subsection{\label{theper} Search for quasi-periodicities}

%----------------------------------------------------

 \begin{figure*}
\centering
\includegraphics[trim=0.cm 0cm 1cm 1.2cm,clip,width=0.74\columnwidth]{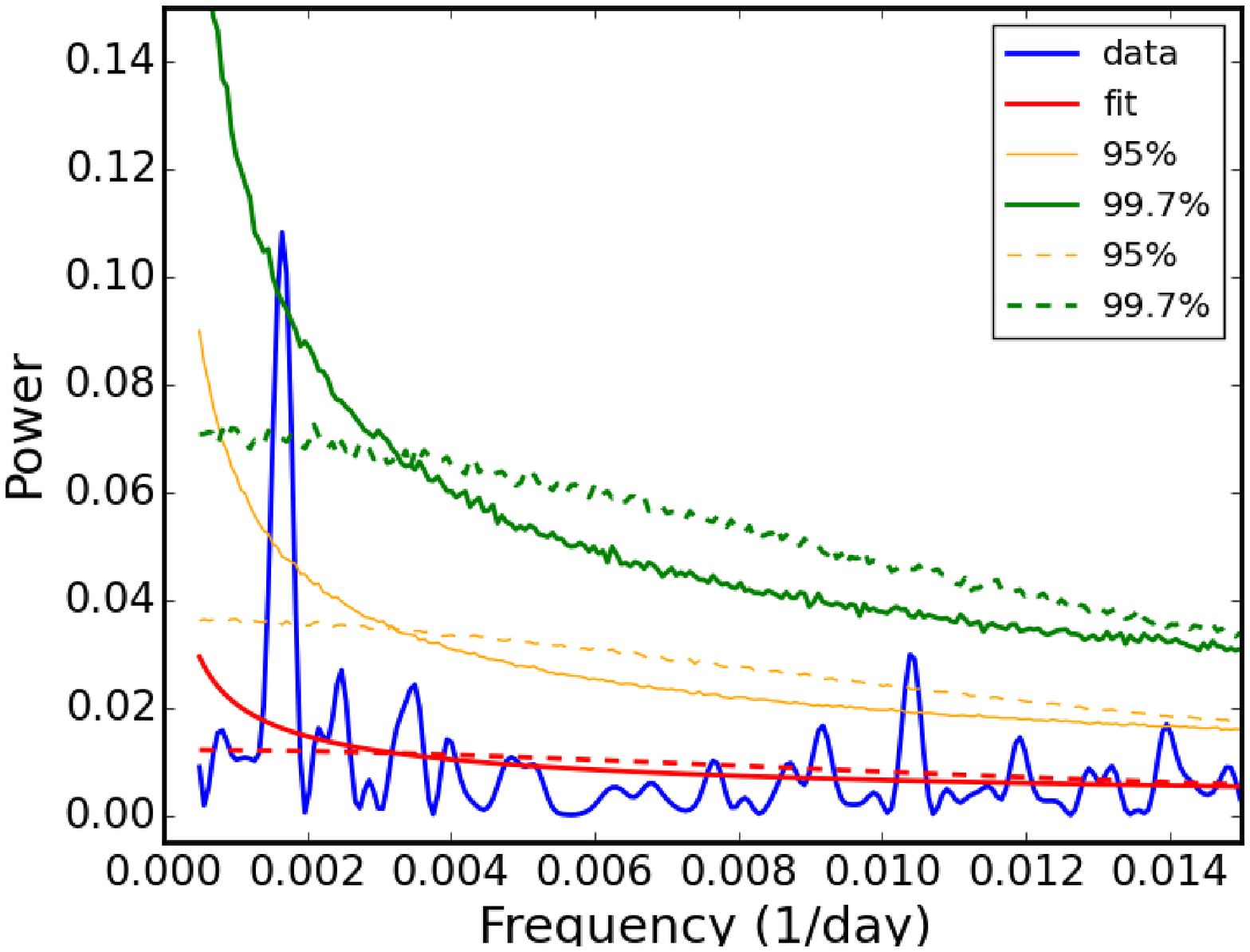}  
\includegraphics[trim=0.cm 0cm 1cm 1.2cm,clip,width=0.74\columnwidth]{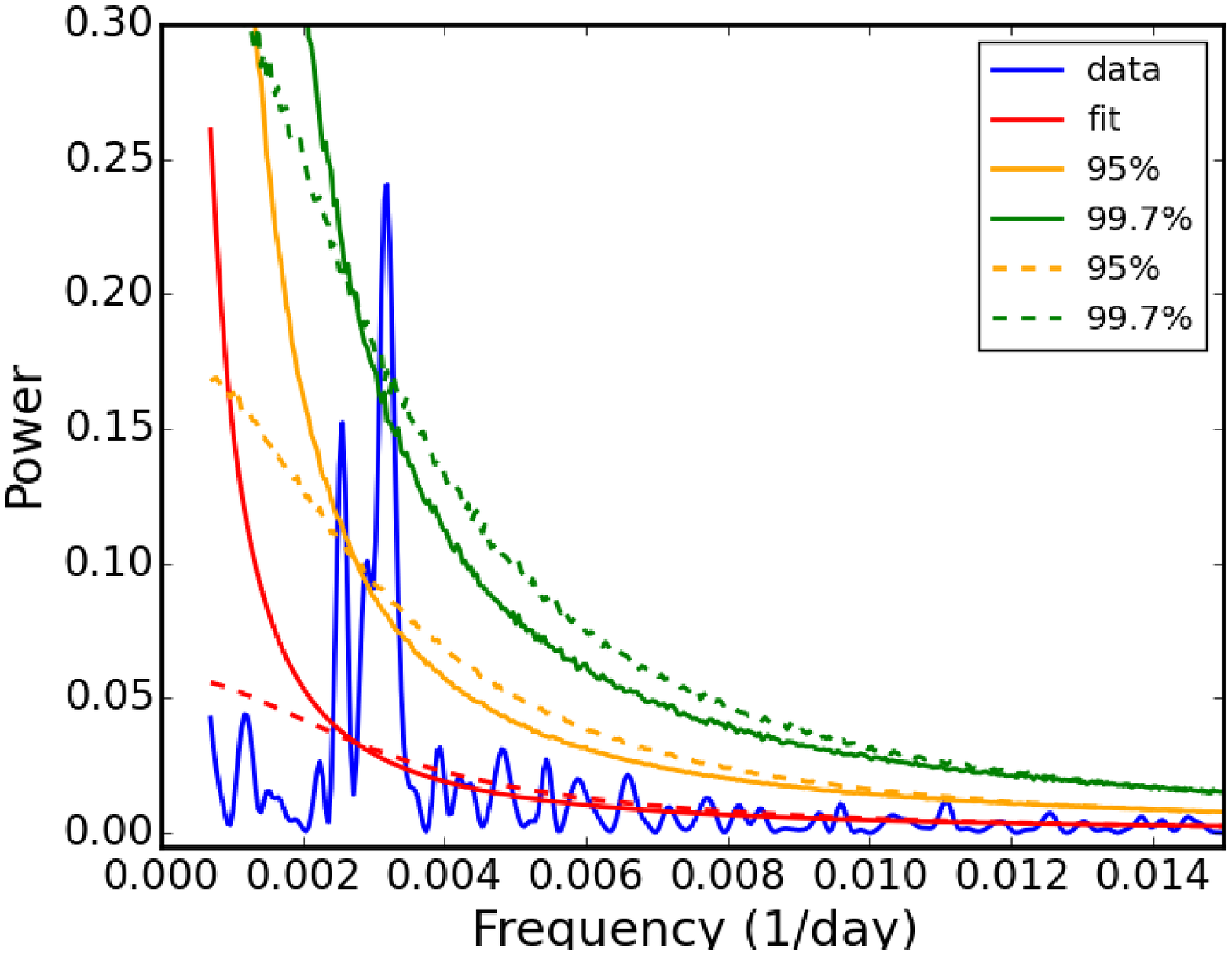}  
\includegraphics[trim=0.cm 0cm 1cm 1.2cm,clip,width=0.74\columnwidth]{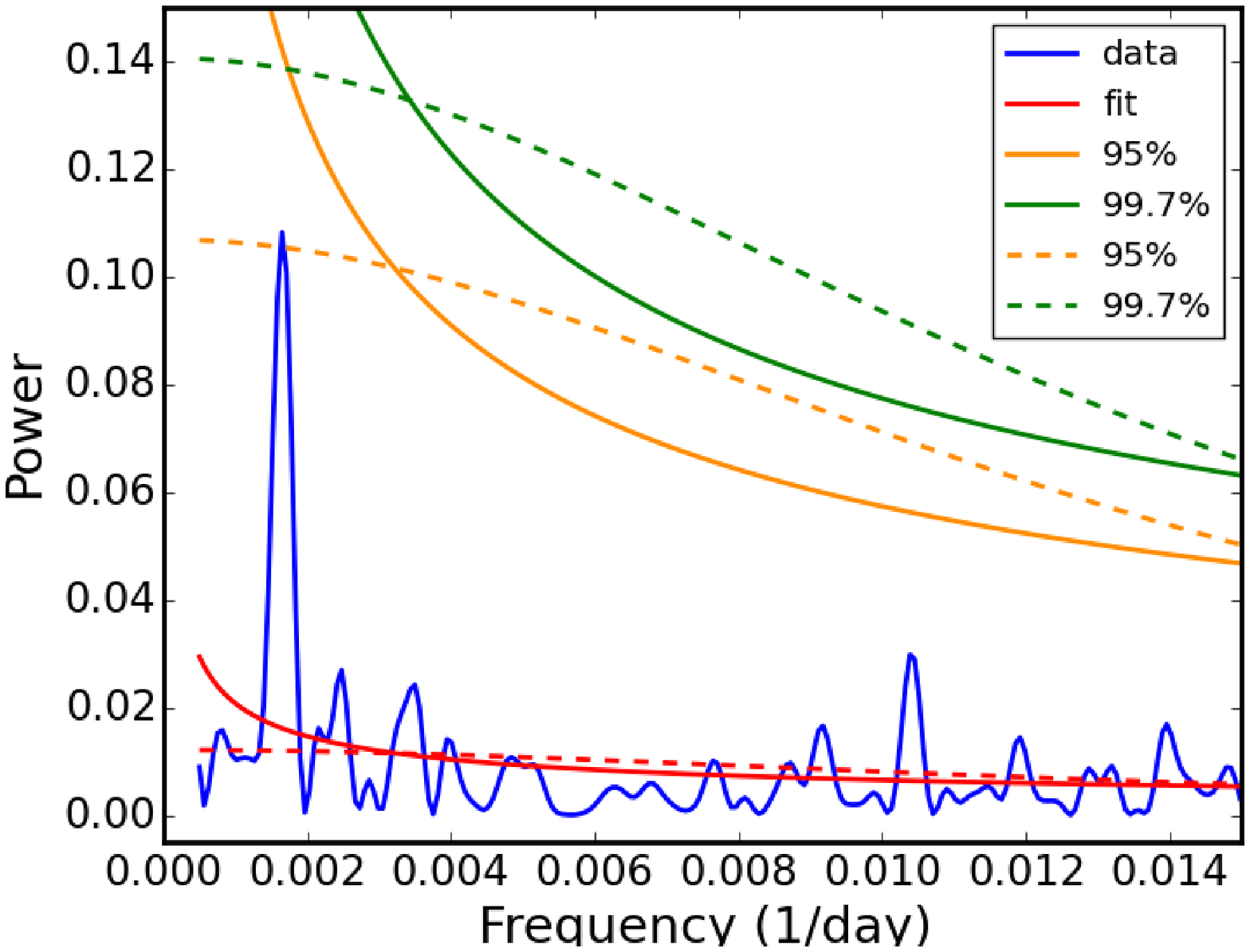}  
\includegraphics[trim=0.cm 0cm 1cm 1.2cm,clip,width=0.74\columnwidth]{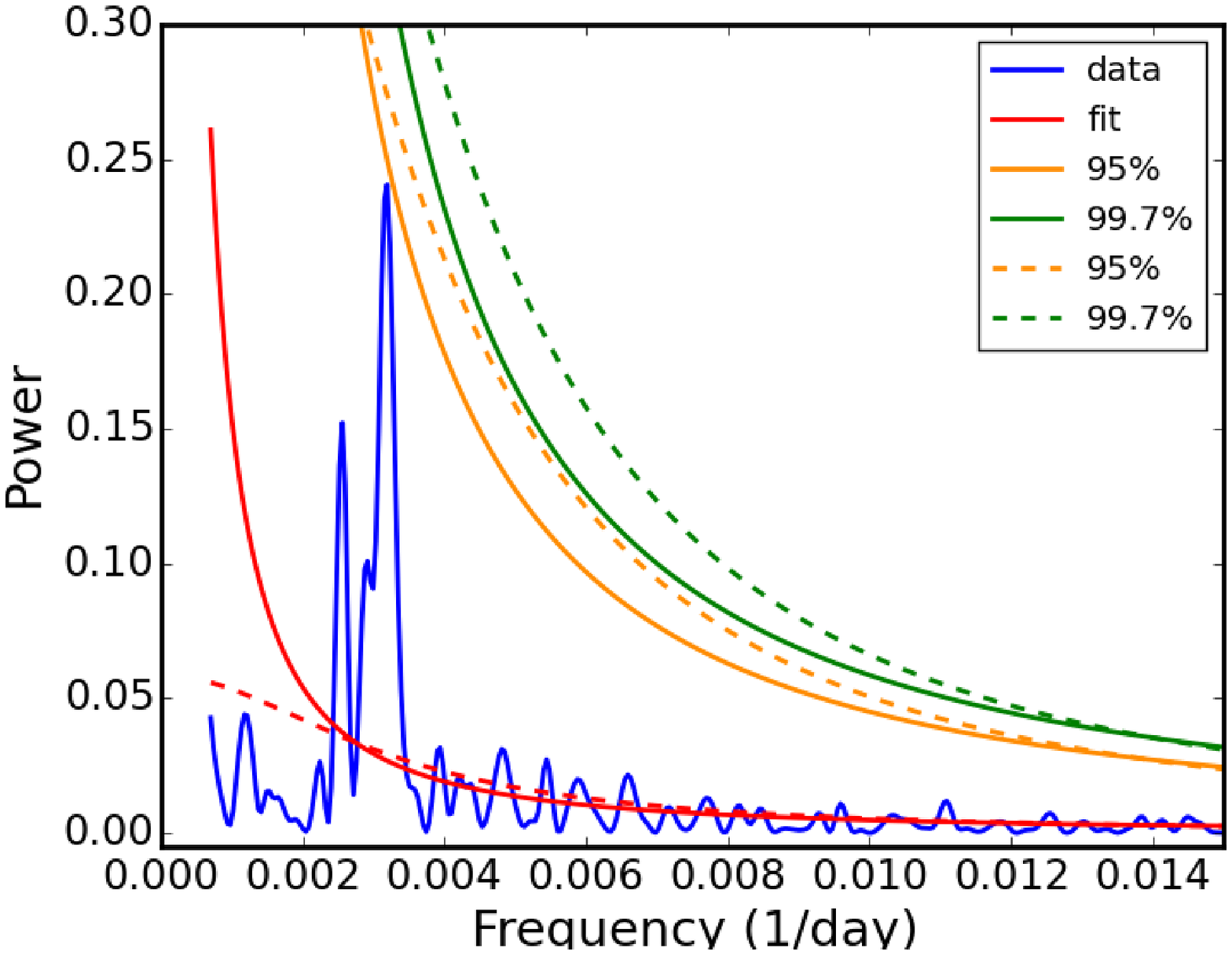}  
\caption{ 
\label{period2155} 
 Single-frequency  \textit{(top panels)} and global-frequency \textit{(bottom panels)} power spectral densities of  PKS2155$-$304.  
\textit{Left panels} are related to the   100 MeV-300 GeV  \textit{Fermi} light curve, and 
 \textit{right panels} to the  R-band light curve.
 Lomb-Scargle spectrum of the input time-series data is given in
 blue and the best-fit   noise spectrum in red.  
The 95.0\% and 99.7\% false alarm levels are reported with yellow and  green lines. 
Solid and dashed lines refer to PL and AR1 models, respectively.    
}  
%%%\end{figure*}  

%%% \begin{figure*}
\centering
\vspace{0.3cm}
\includegraphics[trim=0.cm 0cm 1cm 1.2cm,clip,width=0.74\columnwidth]{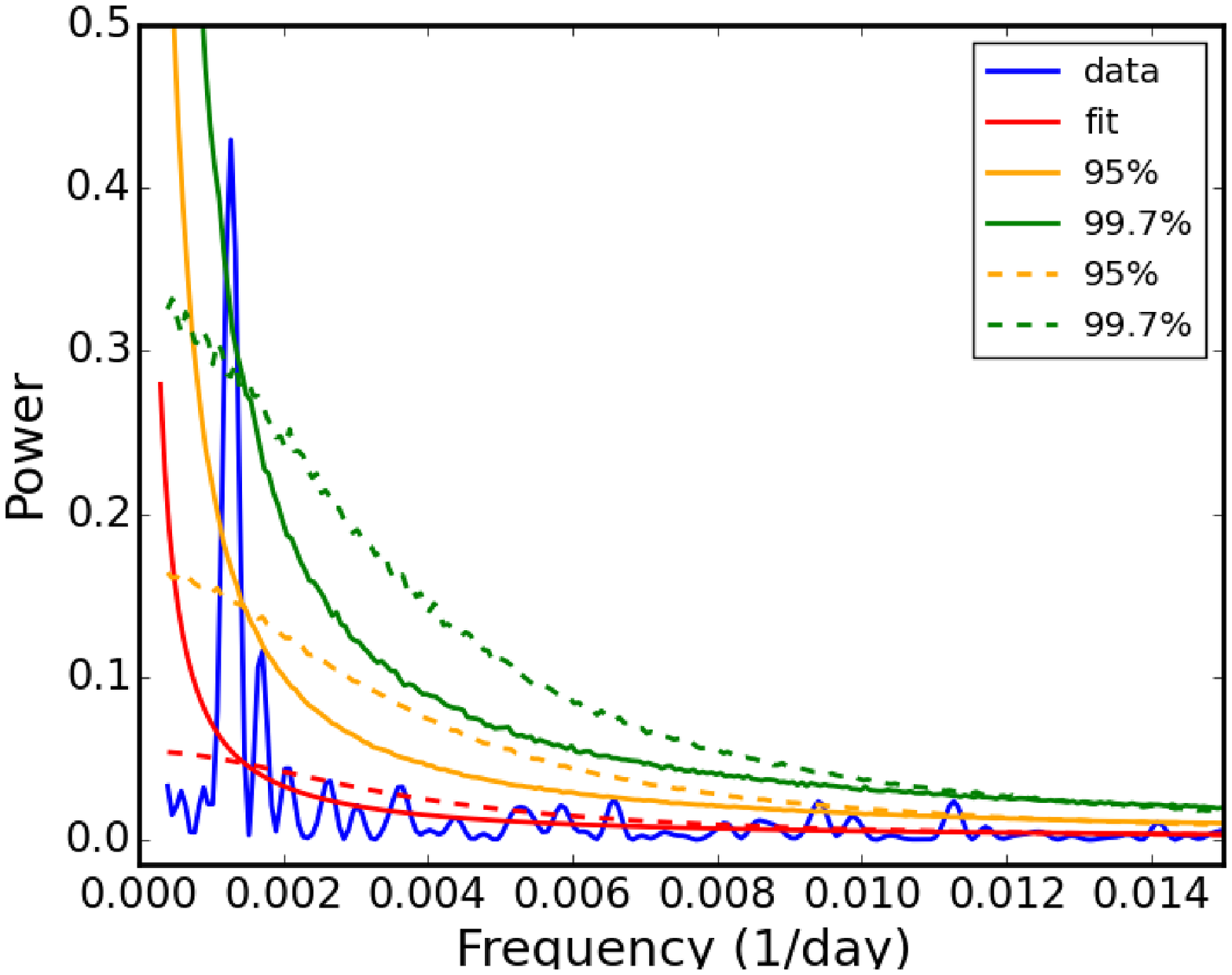}  
\includegraphics[trim=0.cm 0cm 1cm 1.2cm,clip,width=0.74\columnwidth]{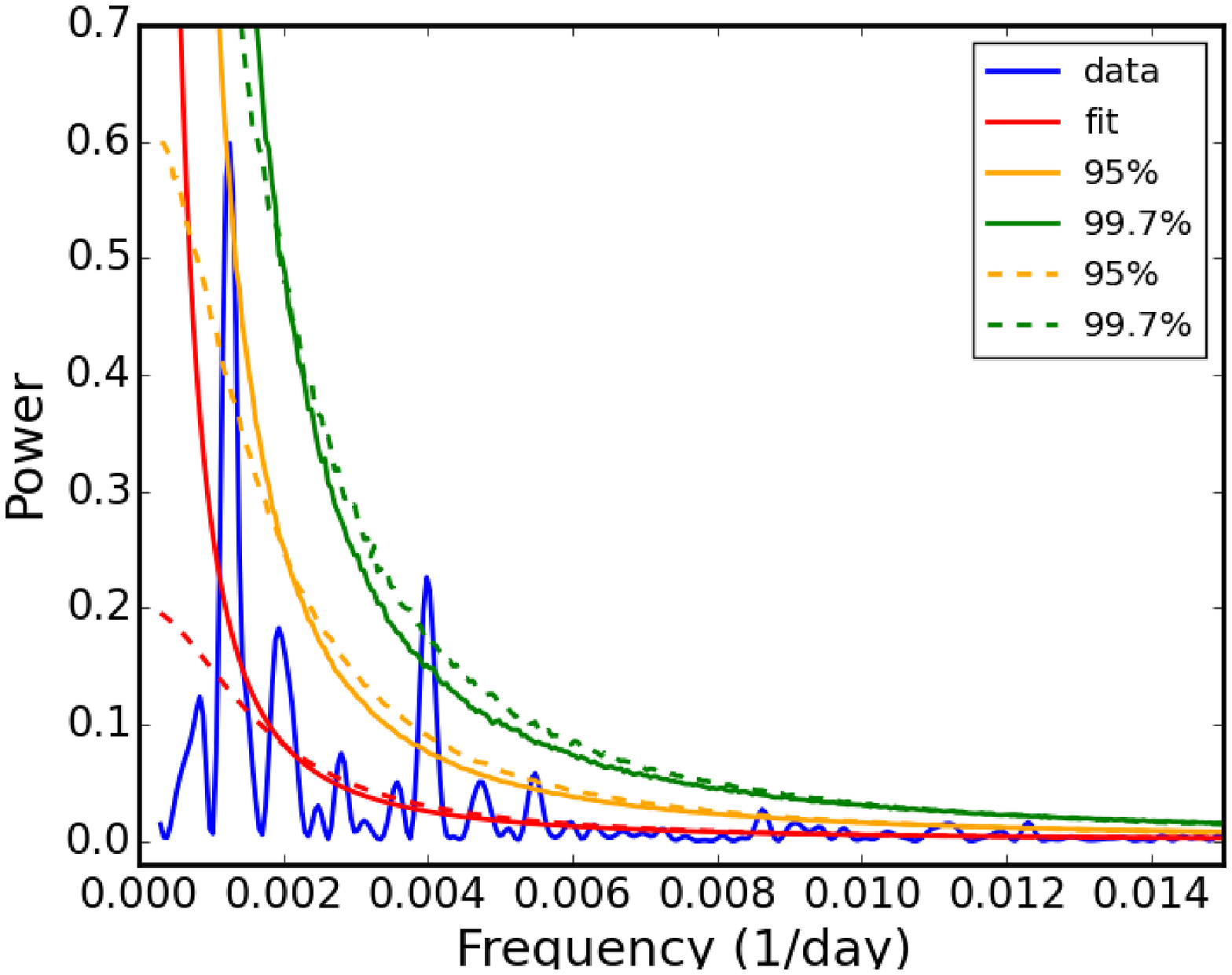}  
\includegraphics[trim=0.cm 0cm 1cm 1.2cm,clip,width=0.74\columnwidth]{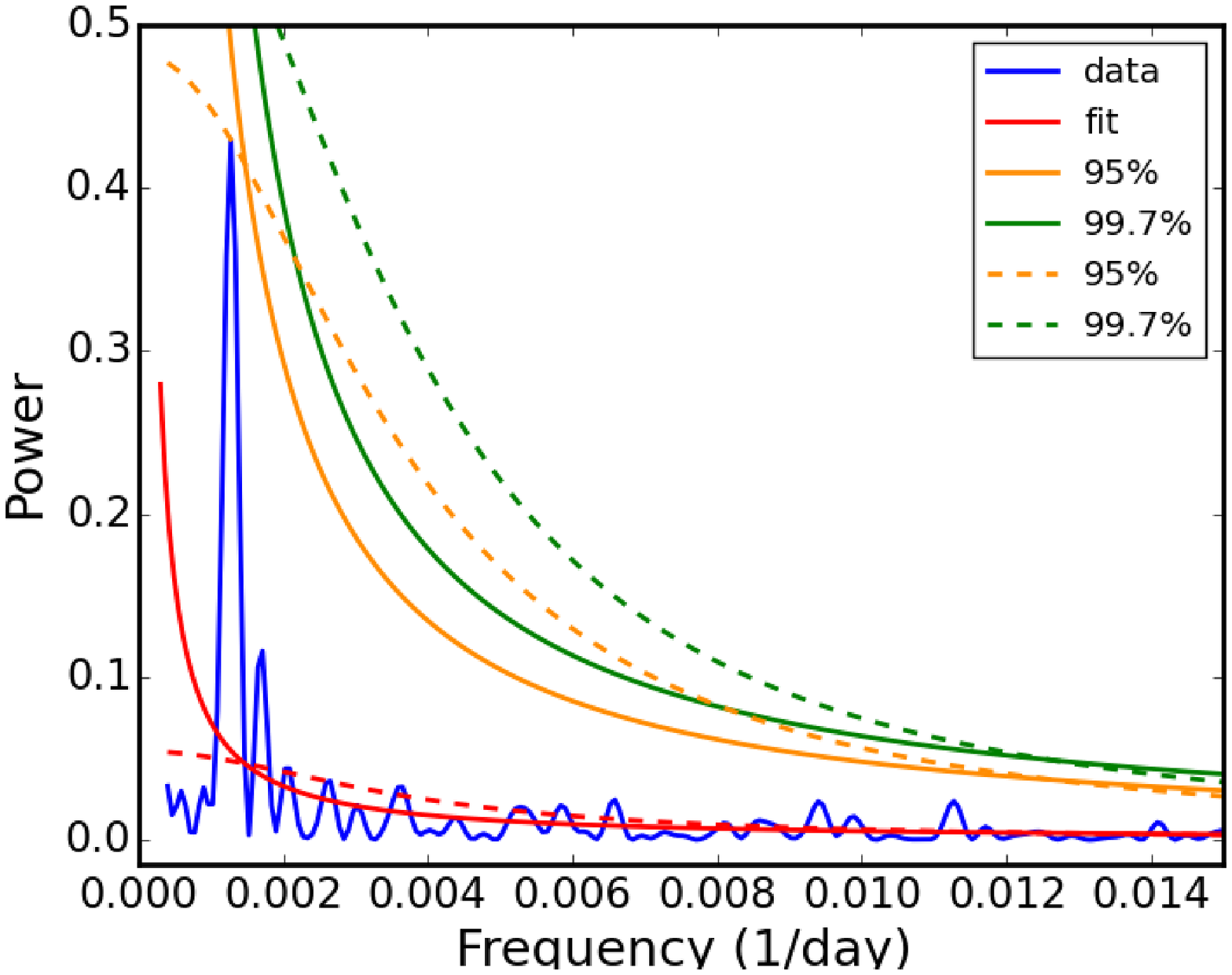}  
\includegraphics[trim=0.cm 0cm 1cm 1.2cm,clip,width=0.74\columnwidth]{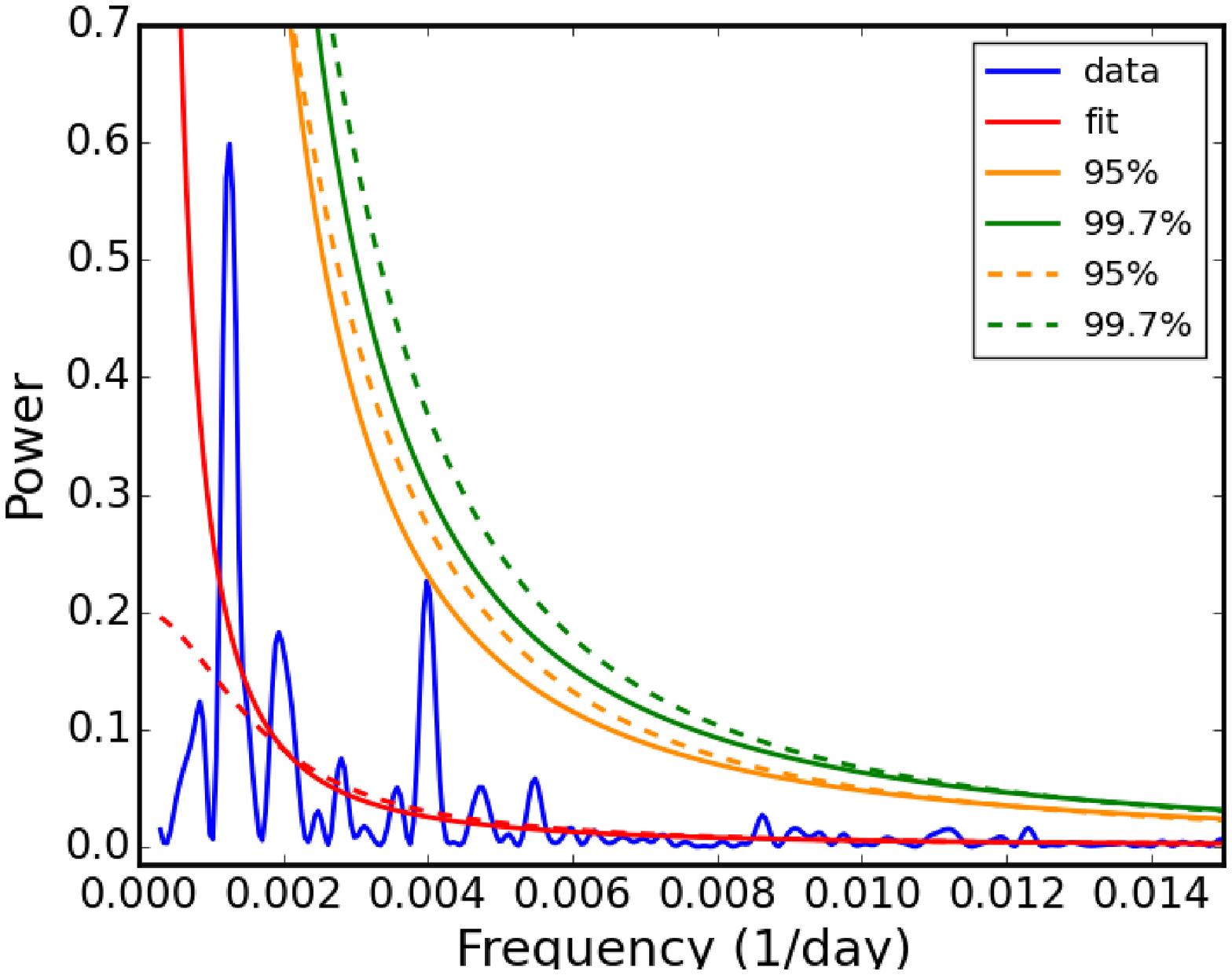}  
\caption{ 
\label{period1553} 
Same as Figure \ref{period2155} for PG1553+113.
}  
\end{figure*}

To search for quasi-periodicities and  to assess their significance against the structured (red) noise
\citep[e.g.][]{Press1978,Vaughan2005}, typically present in blazars, 
we followed  the procedure described by \cite{Vaughan2005,Vaughan2010}, 
\cite{Guidorzi2016}, and \cite {Sandrinelli2017}.
 In short,  we used generalized Lomb-Scargle periodograms \citep[e.g.][]{Scargle1982} 
  to derive the power spectra density (PSD) from the light curves.
Because of their lengths  ($\sim$ 10 to 14 yr), we searched for  periods $\lesssim$ 1000 days,  
and limited the noisy high-frequency periodogram end to periods $\geqslant$ 50 days.
The frequency-dependent noise was modelled as a simple power law (PL) with the form
$S_{\rm PL}(f) = Nf^{-\alpha}$,  
 or as an auto-regression function of the first order (AR1) described as  
 $S_{\rm AR1}(f) = \sigma^2 / (1-2a\cos(2\pi f) + a^2)$, 
where $N$ is the normalization,
$\alpha$ the power-law index, $\sigma$ is the white-noise variance, and $\tau = -1/\ln(a)$ is the 
so-called time constant.
We sampled the best fitting parameters in a Bayesian framework, where  
posterior parameter distributions were obtained using Markov chain Monte  Carlo techniques. 
Simulated PSDs  were derived and the percentiles of the simulated periodograms 
computed at each given frequency (single frequency significances).
Then we evaluated the probability that the power of any peak is equal to or larger than a chosen value
 somewhere in the spectrum 
\citep[global significance, see][for a complete discussion]{Vaughan2010}. 
Optical data for the selected objects (which are all fairly bright sources)
are typically well sampled, and our datasets were binned in order to minimize the possible 
effects of irregular sampling.
Spurious periodicities introduced by the sampling were studied by analysing periodograms 
based on the sampling epochs  \citep[e.g.][]{Vanderplas2017}.
The quality of fits for the PSD modellings were  evaluated using a posterior predictive assessment \citep{Gelman1996}.
The multiple trial correction due to the unknown number of independent sampled frequencies
is included. In fact the same described procedure
    was applied both to the simulated data and to the real data.

%__________________________________________________________________
  
  \subsection{\label{theres} Results}

%__________________________________________________________________

At first we searched for single frequency significance of Lomb-Scargle peaks 
in the optical and \textit{Fermi} light curves  for the two sources, 
PKS2155$-$304 and PG1553+113, applying  PL and AR1 models, as specified in Subsect. \ref{theper}.
Then we computed the global false-alarm probability.
Periodograms are given in Figs. \ref{period2155} and \ref{period1553}  and refer  to the whole length
of the corresponding light curve.

After having modelled the noise with the two PL and AR1  functions,
for both sources there are peaks with modest global frequency significances  (see Table \ref{S}).
However, the examined cases are of interest  because analogous  or harmonic-related 
 results were obtained from studying the optical light-curves.
For PKS2155$-$304, peaks at  T$_{\gamma}\sim$ 620 $\pm$ 41\ days and T$_{opt} \sim$ 315 $\pm$ 25  days
are identified, with T$_{\gamma}$$\sim$ 2 $\cdot$ T$_{opt}$, 
confirming previous results in \cite{Sandrinelli2014a,Sandrinelli2016a}, found by applying different procedures. 
The associated uncertainties were calculated 
by adapting  the mean noise power level method of \cite{Schwar1991} to red-noise spectra. 
The periodograms of PG1553+113 show peaks at the same frequency within the errors, 
corresponding to T$_{\gamma}\sim$ 780 $\pm$ 63 \ days and T$_{opt1} \sim$ 810 $\pm$ 52 days,
similarly to the the quasi-periodic signal found by \cite{Ackermann2015}. 
In this source, higher significances are related to the $\gamma$-ray light curve (see Table \ref{S}).
A  second peak in the optical  at T$_{opt2}  \sim 250 \ \pm\  60$ days is also apparent.
The occurrence of harmonic-related oscillations in the two independent light curves, $\gamma$ and optical, 
in one source, and the concurrence  of the same period in the two bands for the other  one,
although individually modest, suggest  that  the peaks may be related to
a  real quasi-periodicities,  superposed onto chaotic variabilities.

The peaks in the PG1553+113 spectrum at approximately the same
 frequency in the optical and $\gamma$-ray bands appear to be an interesting case,
 which deserves to be further investigated.
To quantify the significance of the detection of the two peaks at the 
same frequency ($T \sim 780-810$ \,days) in the optical and $\gamma$-ray bands,
the  PSDs obtained for the {$\gamma$-ray} and optical data were linearly added and 
 the result was evaluated against a $\chi^2$ distribution with $2M$ degrees of 
 freedom, where $M$ is the number of added PSDs \citep[e.g.][]{Barret2012,Guidorzi2016}. 
 The derived PSD peaks at $T \sim 800$\,days,   but with limited significance, see Table 1.

We note that in the case of PKS2155$-$304 the significances of the optical and $\gamma$-ray 
quasi-periodicities seem to decrease in recent years.
The fading of the major periods may be connected to a decreasing trend of the intensity of the 
source, as  observed in  the case of PKS0537-441 \citep{Sandrinelli2016b}.

Considering also the results for BL Lac \citep[see][and Table \ref{S}]{Sandrinelli2017},
the main conclusion of our analysis is that the three sources reveal
quasi-periodicities with  a moderate yet non-negligible statistical significance. The 
quasi-periodicities occur at the same frequency in both the optical and $\gamma$-rays for BL Lac 
and PG 1553+113, and at correlated frequencies for  PKS2155$-$304.  
We note that the peak significances obtained with the procedure adopted in this paper
 appear modest in comparison with those proposed in the literature and quoted in Section \ref{theintro}.

%__________________________________________________________________

\section{\label{thediscussion}Discussion}

%__________________________________________________________________

In this paper we have reconsidered the search for year-long quasi-periodicities 
in optical and $\gamma$-rays of the two BL Lac sources PKS2155$-$304 and PG1553+113, 
using a procedure adapted from the recipes of \cite{Vaughan2005,Vaughan2010} 
and described in our paper on BL Lac \citep{Sandrinelli2017}. 
 Our results confirm the indication of year-like quasi-periodicities in PKS2155$-$304 and PG1553+113, 
 previously obtained following different numerical schemes \citep{Schulz2002,Ackermann2015}. 
The significances proposed here are lower, because of our
specific consideration of the multi-trial correction.
    
Possible interpretations  of BL Lac objects quasi-periodicities have been proposed by 
\cite{Sandrinelli2014a,Sandrinelli2016a,Sandrinelli2016b,Sandrinelli2017} and by \cite{Ackermann2015}, 
with some  modelling by \cite{Cavaliere2017}, \cite{Sobacchi2017} and \cite{Caproni2017}. 
The suggested pictures can be distinguished into (i) those in which the quasi-periodicities are 
interpreted as due to a SBBH system,  either directly \citep[e.g.][]{Lehto1996,Graham2015}, 
or through some precession process, and (ii) those in which the scenario is instead due to instabilities 
in the relativistic jet \citep[e.g.][]{Camenzind1992,Marscher2014,Raiteri2017}, or in the accretion disk.
The variable significance of  the periodic signal, possibly related to the intensity of the source
(see Section 2.2), or to the noise correlation \citep[e.g.][]{Kelly2014}, does not help us to discriminate 
between the two scenarios.

In this regard the following consideration seems relevant, which starts 
from the estimate of the fraction of possible quasi-periodic objects  
among BL Lacs.  We refer in particular to the recent paper by \cite{Prokhorov2017},
in which a systematic search for cyclical $\gamma$-ray emissions in the \textit{Fermi}-LAT sky is presented.
They examined 7.8 years of data and considered photons  $>$300 MeV 
dividing the sky in 12288 pixels. Searching for periodicities in the 30 days-2.5 years interval in the all sky 
they found seven pixels in which some evidence of periodic signal is present. 
All of these pixels  are identified with a blazar, N$_{BLL}$ = 4 are BL Lac objects: 
PG1553+113,  PKS2155$-$304, BL Lac,  and 0716+714.
For the first three  \cite{Prokhorov2017} confirmed previous quasi-periodic detections 
\citep{Sandrinelli2014a,Ackermann2015,Sandrinelli2016a,Sandrinelli2017}, 
which are also discussed above in this paper.   
The case of 0716+714, which   \cite{Prokhorov2017} mentioned   as a potential
quasi-periodic candidate, was previously examined  in  Figure 3 of \cite{Sandrinelli2017},
who detected the $\gamma$-ray period given by \cite{Prokhorov2017}, but 
noticed the absence of quasi-periodicity in the optical band.

\begin{figure}
\centering
\includegraphics[trim=0cm 0cm 0cm 0cm,clip,width=0.91\columnwidth]{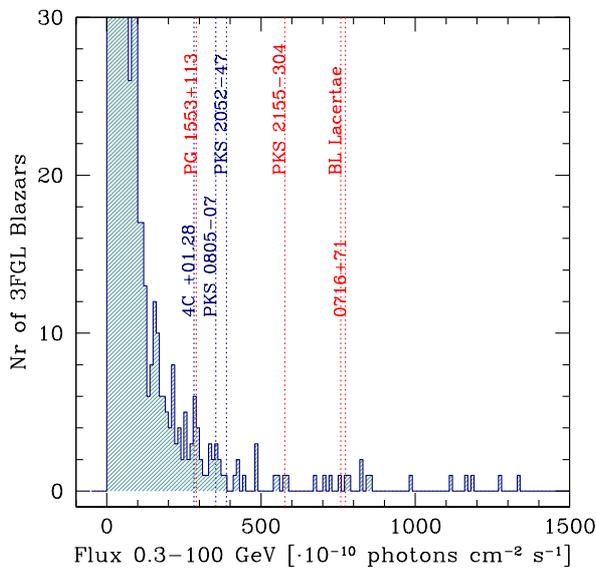}  
\caption{ 
\label{BLLdistr} 
Tail of the distribution of the mean \textit{Fermi} $\gamma$-ray 
 fluxes ($>$300 MeV)  for the 1144 sources classified as blazars
in \textit{Fermi}-LAT 3FGL catalogue  \citep[][]{Acero2015}.
The positions  of the the seven quasi-periodic sources  proposed by 
 \cite{Prokhorov2017} are indicated by dotted  lines, distinguishing 
 BL Lacs (red) and flat spectrum radio quasars (blue).
}
\end{figure}

 In Figure \ref{BLLdistr} we report the distribution of mean \textit{Fermi} $\gamma$-ray 
photon  fluxes ($>$300 MeV)  for the N=1144 sources classified as blazars
in  the \textit{Fermi}-LAT third catalogue  of $\gamma$-ray sources  \citep[3FGL,][]{Acero2015}. 
 It is apparent that a significant detection of possible cyclical behaviour is derived only 
 for the higher flux sources.
  We therefore took   a threshold which corresponds  to the weakest positive  detection 
  of \cite{Prokhorov2017}, f$_{min}$ = 2.8 $\cdot$ 10$^{-8}$
photons cm$^{-2}$ s$^{-1}$, relative to the source 4C +01.28 (3FGL J1058.5+0133), and consider 
the total number N$_{tot}$= 45 of $\gamma$-ray BL Lacs above f$_{min}$/2, which appear 
in the \textit{Fermi}-LAT 3FGL catalogue.  
A rough estimate of the fraction of quasi-periodic  BL Lac objects is therefore given by

\begin{equation}
\label{ }
\xi_B=\frac{N_{BLL}}{N_{tot}} \sim 0.1 
\end{equation}

The value of $\xi_B$ coincides with the estimate already discussed by \cite{Sandrinelli2017}.

\begin{table*}
\begin{center}
\centering
\caption{Significances of the peaks reported in Figs \ref{period2155} and \ref{period1553}. 
Significances $<$ 80\% are indicated with (*).
}
\label{S}
\footnotesize
\begin{tabular}{ccc|cc|cc|cc}
\hline\hline
\multicolumn{3}{c|}{}                                                   &\multicolumn{6}{c}{Significance \ \ \ }\\
\multicolumn{3}{c|}{}                                                   &\multicolumn{2}{c|}{Single-frequency}  &\multicolumn{2}{|c|}{Global-frequency} &\multicolumn{2}{c}{Global-frequency, combined}\\
Source          &T                              &Light curve    &PL                     &AR1                            &PL                     &AR1                            &PL                     &AR1                            \\ 
                        &[days]                 &                       & \%&\%&\%&\% &\%&\%                        \\
\hline
&&&&&&&&\\
PG1553+113      &780 $\pm$ 63   &100 MeV-300 GeV        &99.97          &99.97          &90             &95     &80&80\\                
                        &810 $\pm$ 52   &R                              &95                     &99                     &(*)                    &(*)            &&\\                
&&&&&&&&\\
PKS2155$-$304   &620 $\pm$ 41   &100 MeV-300 GeV        &99.9           &99.99                          &(*)                         &95             &&\\
                        &315 $\pm$ 25 &R                                &99.99          &99.97                          &95                     &90             &&\\
&&&&&&&&\\
BL Lac          &680 $\pm$ 35 &100 MeV-300 GeV  &95                     &99.7                           &(*)                    &(*)            &95     &99.7   \\
                        &670 $\pm$ 40 &R                                &99.5           &99.9                           &80                     &80             &&   \\
&&&&&&&&\\
\hline
\end{tabular}
\end{center}
\end{table*}

We now compare the above with results of a similar discussion of optical quasi-periodicities of quasars.
In the 250k quasars of the Catalina Real-time Transient Survey 
there are 111 potentially periodic sources \citep{Graham2015},
and in the Palomar Transient Factory  \citep[][]{Charisi2016} in a sample of 33k objects
 the possible quasi-periodicities are 33,  yielding $\xi_Q$$\sim$ 4$\cdot$10$^{-4}$ - 10$^{-3}$. 
Let us restrict to the local universe z$\lesssim$1.
The quasar density $\rho_Q$ $\sim$ 10$^{-7}$ Mpc$^{-3}$ \citep[e.g.][]{Croom2004}, while the BL Lac  
density from the\textit{ Fermi} $\gamma$-ray data \citep{Ajello2014} is $\rho_B$ $\sim$10$^{-8}$ Mpc$^{-3}$.  
The density of quasi-periodic BL Lacs is given by $\rho_{BP}$ = $\xi_B$ $\rho_B$$\sim$10$^{-9}$ Mpc$^{-3}$, 
and that of quasi-periodic quasars by $\rho_{QP}$ = $\xi_Q$  $\rho_Q$$\sim$ 10$^{-11}$-10$^{-10}$ Mpc$^{-3}$. 
Then at face value $\rho_{BP}$$>>$$\rho_{QP}$;  the local density quasi-periodic 
BL Lacs largely overcomes that of quasars.
We stress again that the argument is based on a search
of quasi-periodicities for BL Las and quasars in two very different spectral bands.
      
The higher fraction of quasi-periodic sources for BL Lacs with respect to quasars 
might have important consequences for its interpretation. In fact \cite{Sesana2018} 
 have shown that if all potentially periodic quasars 
were SBBHs, with some assumption about the mass ratio of the two black holes 
there would be a tension with the gravitational wave background at  nanoHz frequencies
 (year long) probed with  pulsar timing array. 
 This tension would be likely substantially increased in the case of BL Lacs objects.  
This subject is treated in detail in a coordinated paper of ours \citep{Holgado2018} 
 in which we will indeed show that 
this is the case. It is therefore likely that BL Lac quasi-periodicities 
are not related to a binary nature of the sources; they can be caused by 
other physical processes such as relativistic jet instabilities, or they can still 
manifest as chance fluctuation of the source stochastic variability
since the significance of the quasi-periodicities is noticeable ($\sim$90\%), 
but not extreme.

\begin{acknowledgements}
We thank the referee for the prompt and constructive report. 
A. Sesana is supported by the Royal Society.
\end{acknowledgements}

\end{document}